\documentclass[%
 aip,
 amsmath,amssymb,
 preprint,%
]{revtex4-2}

\usepackage{graphicx}
\usepackage{dcolumn}
\usepackage{bm}

\usepackage[T1]{fontenc}
\usepackage{mathptmx}
\usepackage{siunitx}
\DeclareSIUnit\Oersted{Oe}
\DeclareSIUnit\electronvolt{eV}
\DeclareSIUnit\emu{emu}
\DeclareSIUnit\atomp{at.\%}
\usepackage{xcolor}
\graphicspath{{./Figure-review/}}

\begin{document}


\title[Highly fcc-textured Pt-Al alloy films grown on MgO(001) showing enhanced spin Hall efficiency]{Highly fcc-textured Pt-Al alloy films grown on MgO(001) showing enhanced spin Hall efficiency}

\author{Yong-Chang Lau}
\email{lau.yong.chang.d8@tohoku.ac.jp}
\affiliation{Institute for Materials Research, Tohoku University, Sendai 980-8577, Japan}
\affiliation{Center for Spintronics Research Network, Tohoku University, Sendai 980-8577, Japan}
\author{Takeshi Seki}
\affiliation{Institute for Materials Research, Tohoku University, Sendai 980-8577, Japan}
\affiliation{Center for Spintronics Research Network, Tohoku University, Sendai 980-8577, Japan}
\author{Koki Takanashi}
\affiliation{Institute for Materials Research, Tohoku University, Sendai 980-8577, Japan}
\affiliation{Center for Spintronics Research Network, Tohoku University, Sendai 980-8577, Japan}
\affiliation{Center for Science and Innovation in Spintronics, Core Research Cluster, Tohoku University, Sendai 980-8577, Japan}

\date{\today}

\begin{abstract}
We report on a systematic comparative study of the spin Hall efficiency between highly face-centered cubic (fcc)-textured Pt-Al alloy films grown on MgO(001) and poorly-crystallized Pt-Al alloy films grown on SiO$_2$. Using CoFeB as the detector, we show that for Al compositions centering around $x = 25$, mainly L1$_{2}$ ordered Pt$_{100-x}$Al$_x$ alloy films grown on MgO exhibit outstanding charge-spin conversion efficiency. For Pt$_{78}$Al$_{22}$/CoFeB bilayer on MgO, we obtain damping-like spin Hall efficiency as high as $\xi_\textrm{DL} \sim +0.20$ and expect up to seven-fold reduction of power consumption compared to the polycrystalline bilayer of the same Al composition on SiO$_2$. This work demonstrates that improving the crystallinity of fcc Pt-based alloys is a crucial step for achieving large spin Hall efficiency and low power consumption in this material class.

\end{abstract}

\maketitle

\section{\label{sec:Intro}Introduction}

Current-induced spin-orbit torque (SOT) \cite{Manchon_SOTreview} is a promising means for manipulating the magnetization of a nanomagnet \cite{Miron2011,Liu2012,Yu2014,FukamiPtMn2016,Lau2016,Avci2017,DC2018,Khang2018,Baek2018,Sato2018,WangNiO2019,WangFGT2019,Grimaldi2020,Peng2020,Liu2021,dc2020} and for developing next-generation magnetic memories \cite{Garello2019}. In a non-magnetic material (NM)/ferromagnetic material (FM) bilayer heterostructure with strong spin-orbit coupling, application of an in-plane charge current leads to the generation of a transverse spin current and accumulation of non-equilibrium spin density near the NM/FM interface via either the "bulk" spin Hall effect (SHE) \cite{Sinova_SHEreview} or interfacial Rashba-Edelstein effect \cite{Sanchez2013,Vaz2019} or spin-momentum locking of the topological surface states \cite{Mellnik2014}. The accumulated spin can be absorbed by the FM, exerting damping-like and field-like SOT to the magnetization. Experimentally, this charge-to-spin conversion process is commonly expressed by the relation:
\begin{eqnarray}
\label{eq:SHA}
j_\mathrm{spin,FM}^\mathrm{DL(FL)}=\frac{\hbar}{2e}\xi_\mathrm{DL(FL)}j_\mathrm{charge,NM}
\end{eqnarray}
where $\hbar$ is the reduced Planck constant, $e$ the elementary charge, $j_\mathrm{charge,NM}$ the charge current flowing within the NM layer, $j_\mathrm{spin,FM}^\mathrm{DL(FL)}$ the equivalent spin current absorbed by the FM layer for producing the measured damping-like (field-like) SOT, and $\xi_\mathrm{DL(FL)}$ the damping-like (field-like) spin Hall efficiency. Note that here $\xi_\mathrm{DL(FL)}$ describes phenomenologically the conversion efficiency based on the total spin current eventually absorbed by the FM for SOT generation, ignoring its origin (e.g. SHE or other interfacial mechanisms) and its transmission probabilities across the interface (e.g. spin backflow and spin memory loss) \cite{Bass_2007,Haney2013,Sanchez_PRL_SML,Zhang2015,Pai2015,Zhu_PRLSOC}. Concerning the power consumption, the figure of merit scales with $\xi_\textrm{DL}^2\rho_\textrm{NM}$ or $\xi_\textrm{DL}\sigma_\textrm{SH}$ where $\rho_\textrm{NM}$ is the longitudinal resistivity and $\sigma_\textrm{SH}$ the spin Hall conductivity of the NM of thickness $t_\textrm{NM}$. More rigorously, for NM/FM bilayer relevant for most applications, one should include the power dissipation due to the unavoidable current flow within the FM layer (with a thickness $t_\textrm{FM}$ and resistivity $\rho_\textrm{FM}$), leading to the following power efficiency parameter $\eta \equiv \frac{1}{1+s}\xi_\textrm{DL}\sigma_\textrm{SH}$ with $s = t_\textrm{FM}\rho_\textrm{NM}/(t_\textrm{NM}\rho_\textrm{FM})$. Finding material combinations that maximize $\eta$ is of paramount importance for improving the performance and competitiveness of any spintronic technology involving SOT \cite{Jiang2015,Zahed2020,Luo2020,Song2020}.

Pt is an archetypal NM for efficient charge-to-spin conversion and is well-known for being the elemental material exhibiting the largest $\sigma_\textrm{SH}$. Referring to $\eta$, however, elemental Pt with relatively small $\xi_\textrm{DL}$ is less attractive. Meanwhile, the large $\sigma_\textrm{SH}$ of Pt is mainly attributed to the intrinsic Berry curvature mechanism \cite{Guo_PRL_PtSHC,Sagasta_PRB2016}, featuring $\sigma_\textrm{SH}$ that is independent of the carrier relaxation time, $\tau \propto 1/\rho_\textrm{NM}$ for the conduction in the "moderately dirty" regime. An immediate strategy for improving $\eta$ would be to reduce $\tau$ while maintaining the high $\sigma_\textrm{SH}$ by alloying \cite{Lowitzer_PRL_alloytheory}. Following the pioneering demonstration of enhanced $\xi_\textrm{DL}$ and $\eta$ in polycrystalline Pt-Al and Pt-Hf alloys \cite{PtAl_APL2016}, considerable research efforts have been devoted to investigate the SOT in Pt-based alloys \cite{Obstbaum_PRL2016PtAu,Zhu-PtAu,Zhu_PtPd,Zhu_PtMgO,Shu_PRMaterials_CuPt,Tian2019_CuPt,Hu_Pai_CuPt,Seki2019,Quan_2020_PtCr,PtMn_alloy_APL}. It is worth pointing out that among these attempts, alloying Pt with an element of identical face-centered cubic (fcc) crystal structure such as Au \cite{Obstbaum_PRL2016PtAu,Zhu-PtAu}, Pd \cite{Zhu_PtPd}, and Cu \cite{Shu_PRMaterials_CuPt,Tian2019_CuPt,Hu_Pai_CuPt} tends to maintain or even slightly enhance $\sigma_\textrm{SH}$ over an extended alloying concentration of $x \gtrsim 25$~at.\%. Intercalating thin Ti spacers in Pt to form a fcc(111)-textured multilayer is also effective for maintaining high $\sigma_\textrm{SH}$ while reducing $\tau$ (Ref.~\cite{Zhu_PtTi_multi}). In contrast, introducing dopants of different crystal structures and with limited solubility (e.g. Al, Hf \cite{PtAl_APL2016}, MgO \cite{Zhu_PtMgO}, Cr \cite{Quan_2020_PtCr}, and Mn \cite{PtMn_alloy_APL}) in Pt typically results in a rapid reduction of $\sigma_\textrm{SH}$ and deterioration of the fcc lattice. These observations seem to suggest increasing the resistivity while maintaining the fcc crystal structure is beneficial for achieving large $\eta$ in Pt-based alloys. Here, by revisiting the charge-to-spin conversion in Pt-Al binary alloys, we show that highly fcc-textured Pt-Al alloy films grown on MgO(001) single crystal substrates exhibit up to seven-fold enhancement of $\eta$ compared to their poorly-crystallized polycrystalline counterparts of the same composition grown on SiO$_2$ substrates. Our results highlight the central role of high fcc crystallinity and atomic ordering that govern the efficiency of charge-to-spin conversion in these binary alloys.

\section{\label{sec:Method}Experimental Methods}
\begin{figure}
\begin{center}
  \includegraphics
  [width=0.45\columnwidth]
  {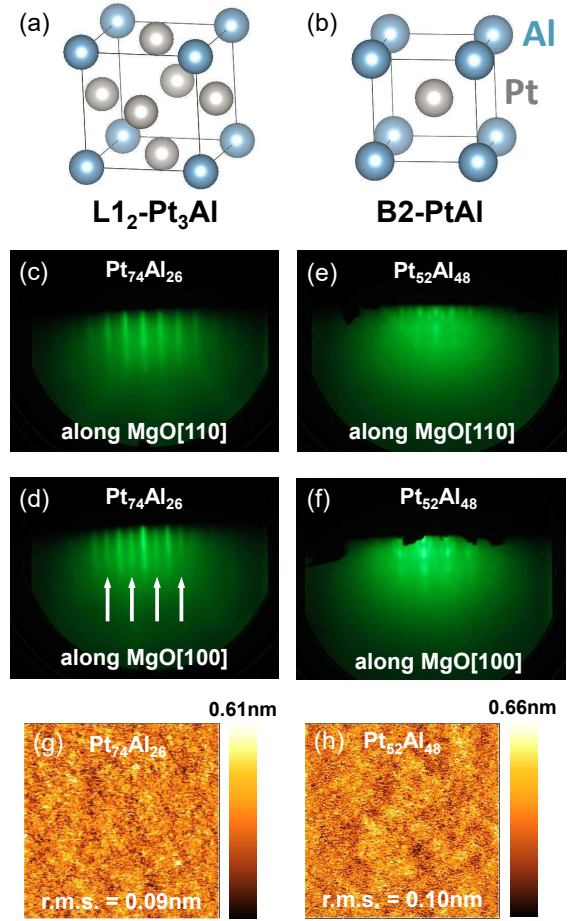}
  \caption{
  (a-b) Illustrations of the crystal structure for (a) face-centered cubic (fcc) L1$_2$-ordered Pt$_3$Al and (b) body-centered cubic B2-ordered PtAl. (c-f) Reflection high-energy electron diffraction (RHEED) patterns of (c-d) Pt$_{74}$Al$_{26}$ and (e,f) Pt$_{52}$Al$_{48}$ films grown on MgO(001) substrates. (c, e) and (d, f) were observed with the electron beam along the MgO[110] and MgO[100] azimuths, respectively. White arrows in (d) indicate superlattice streaks due to the L1$_2$ ordering. (g-h) $\SI{1}{\micro\meter} \times \SI{1}{\micro\meter}$ atomic force microscopy images of uncapped (g) Pt$_{74}$Al$_{26}$ and (h) Pt$_{52}$Al$_{48}$ films showing low root mean square (r.m.s.) roughness of $\sim \SI{0.1}{\nano\meter}$.}
  \label{fig:RHEED}
\end{center}
\end{figure}

Thin film heterostructures consisting of MgO(001) or thermally-oxidized Si/SiO$_2$ (substrate)//\allowbreak Pt$_{100-x}$Al$_x$($\sim$6)\allowbreak/Co$_{20}$Fe$_{60}$B$_{20}$(2)/Al(3) (thicknesses in nanometer) were grown using an ultra-high vacuum magnetron sputtering tool with a base pressure better than \SI{2E-7}{\pascal}. Pt$_{100-x}$Al$_x$ alloy films of varying Al concentration $x$ ranging from 0 to $\sim 48$~at.\% were obtained by tuning the sputtering power of elemental Pt and Al targets. The growth rate of these Pt-Al alloys was around \SI{0.03}{\nano\meter\per\second}. Al was used as the capping layer which naturally oxidized into AlO$_x$. Deposition on MgO(001) was carried out by first heating the blank substrate to a substrate temperature $T_\textrm{sub} = \SI{500}{\celsius}$ for $\sim$ 1 hour to remove the magnesium hydroxide and other contaminations on the surface. Next Pt$_{100-x}$Al$_x$ was deposited at $T_\textrm{sub} = \SI{300}{\celsius}$ followed by postannealing at the same $T_\textrm{sub}$ for $\sim\SI{45}{\minute}$. The remaining CoFeB/Al layers were grown after cooling down the substrate to near ambient temperature. Prior to this deposition, reflection high-energy electron diffraction (RHEED) with a beam energy of \SI{20}{\kilo\volt} was observed for the Pt$_{100-x}$Al$_x$ surface of selected samples. The full stacks on SiO$_2$ were also grown at ambient temperature. Grazing-angle x-ray reflectivity (XRR) was used to measure the thickness and estimate the density of Pt$_{100-x}$Al$_x$. The Al concentration $x$ of the films were estimated by extrapolating the obtained x-ray density to those of the three known compounds: Pt (Density = 21.45), L1$_2$ Pt$_3$Al (Density = 17.58) and B2 PtAl (Density = 13.13). From the XRR, we found no noticeable change of the growth rate for the two sample series, albeit with different substrates and $T_\textrm{sub}$ (See the supplementary material). Out-of-plane and in-plane x-ray diffraction (XRD) were measured to study the structure and the epitaxial relationship of Pt$_{100-x}$Al$_x$ with respect to the substrate. CoFeB and Al are practically amorphous showing negligible contribution to the XRD spectra. Surface morphology was characterized by atomic force microscopy (AFM) on Pt$_{100-x}$Al$_x$ films without CoFeB and Al capping. Magnetic properties of Pt$_{100-x}$Al$_x$/CoFeB bilayers were measured using a vibrating sample magnetometer (VSM) at room temperature.

Micron-sized Hall bar devices were fabricated using standard optical lithography and Ar ion milling. The distance between the two longitudinal voltage probes $L$ and the width of the Hall bar $w$ are 25 and \SI{10}{\micro\meter}, respectively. Cr(10)/Au(100) contact pads were formed using ion beam sputtering and standard lift-off process. The variable-temperature magnetotransport was measured in a Quantum Design Physical Property Measurement System (PPMS) equipped with a horizontal rotator using the built-in resistivity bridge. Harmonic Hall measurements were also carried out in the PPMS at \SI{300}{\kelvin} via a home-made external measurement platform for quantifying the spin Hall efficiency. A sinusoidal current of frequency \SI{172.1}{\hertz} was applied using a Keithley 6221 dc and ac current source meter while the first and second harmonic Hall resistances were simultaneously measured using two LI5660 lock-in amplifiers.

\section{\label{sec:Results}Experimental Results}

\subsection{\label{subsec:Structure}Structural characterizations}
The binary Pt-Al phase diagram for bulk \cite{Okamoto} suggests that the solubility of Al in fcc-Pt is merely $\sim$ 10 at.\%. On increasing the Al concentration, we encounter several ordered Pt-Al compounds that crystallize in cubic structures. We focus on fcc-based L1$_2$ Pt$_3$Al and body-centered cubic (bcc)-based B2 PtAl, for which the structures are illustrated in Fig.~\ref{fig:RHEED}a and b, respectively. Of particular interest is the fcc-based L1$_2$ Pt$_3$Al. Based on the arguments elaborated in Sec.~\ref{sec:Intro}, the formation of this fcc-ordered compound may help to maintain a higher $\sigma_\textrm{SH}$, which is a main motivation of this study.

\begin{figure*}
\begin{center}
  \includegraphics[width=0.9\columnwidth]{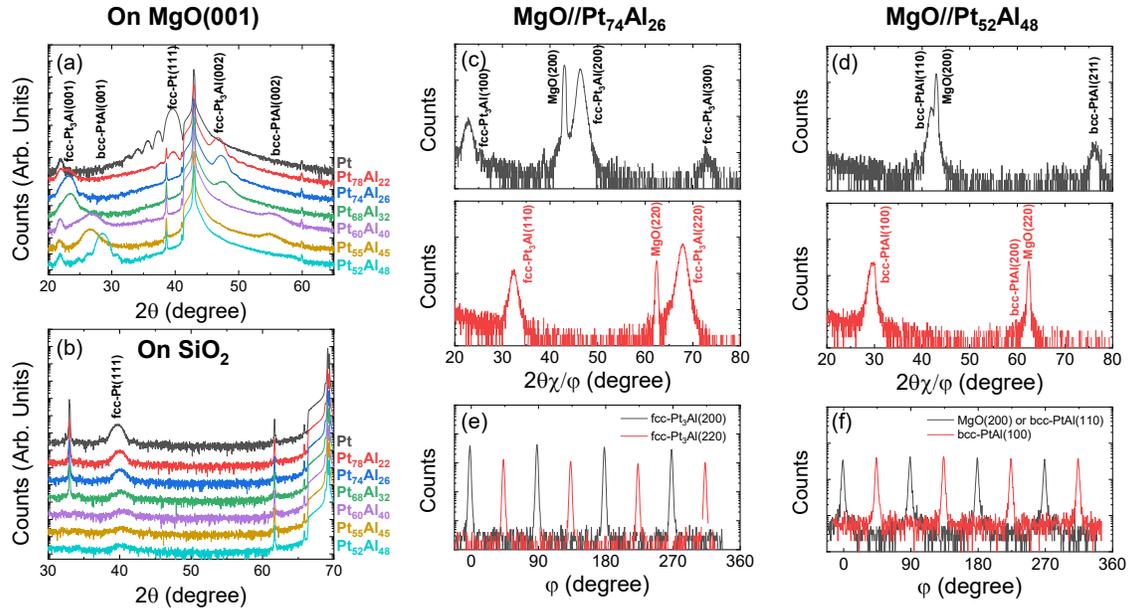}
  \caption{
  (a-b) Out-of-plane 2$\theta$-$\theta$ x-ray diffraction (XRD) spectra of Pt$_{100-x}$Al$_x$ films with varying Al concentration $x$, grown on (a) MgO(001) and (b) thermally oxidized Si substrates. (c-d) In-plane $2\theta\chi/\varphi$ scans of (c) Pt$_{74}$Al$_{26}$ and (d) Pt$_{52}$Al$_{48}$ films on MgO (i.e. MgO//Pt$_{74}$Al$_{26}$ and MgO//Pt$_{52}$Al$_{48}$), probing lattice planes along MgO[100] (upper panels) and MgO[110] (lower panels). (e-f) In-plane $\varphi$ scans showing the four-fold symmetry of the films and the MgO substrate. The thicknesses of these Pt$_{100-x}$Al$_x$ films are approximately \SI{6}{\nano\meter}.}
  \label{fig:XRD}
\end{center}
\end{figure*}

We first present the successful growth of epitaxial, nearly-stoichiometric L1$_2$-ordered Pt$_{74}$Al$_{26}$. Figure \ref{fig:RHEED}c and d show the RHEED patterns of the Pt$_{74}$Al$_{26}$ film with electron beam along the MgO[110] and MgO[100]. The clear streak patterns that systematically change with the substrate rotation imply that the film was epitaxially grown on MgO with a flat surface. Furthermore, the appearance of streaky satellite pattern (indicated by white arrows) along the MgO[100] is consistent with L1$_2$ ordering. In contrast, the corresponding RHEED patterns for Pt$_{52}$Al$_{48}$, depicted in Fig.~\ref{fig:RHEED}e and f, are more complicated, suggesting a mixture of several crystal orientations. AFM images of these two films are compared in Fig.~\ref{fig:RHEED}g and h. Both films exhibit very low surface roughness with a root-mean-square (r.m.s.) roughness of the order of \SI{0.1}{\nano\meter}. The flat surface morphology cannot explain the spot-like RHEED patterns for Pt$_{52}$Al$_{48}$.

\begin{table*}
\caption{Density, out-of-plane lattice parameter $c$, in-plane lattice parameter $a$, and tetragonal distortion $c/a$ extracted from the x-ray measurements for $\sim\SI{6}{\nano\meter}$ L1$_2$ Pt$_{74}$Al$_{26}$ and B2 Pt$_{52}$Al$_{48}$ films grown on MgO. $a$ is compared with that of MgO. The mismatch in \% and the strain are also given.}
\label{tab:XRD_summary}
\begin{ruledtabular}
\begin{tabular}{ccccccc}
Compound&Density&Out-of-plane, $c$ ({\AA})&In-plane, $a$ ({\AA})&$c/a$&In-plane MgO ({\AA})&Mismatch\\
\hline
L1$_2$ Pt$_{74}$Al$_{26}$ & 17.4 & 3.84 & 3.91 & 0.98 & $a_\textrm{MgO}$ = 4.21 & 7\% tensile\\
B2 Pt$_{52}$Al$_{48}$ & 13.5 & 3.12 & 3.01 & 1.04 & $a_\textrm{MgO}/\sqrt{2}$ = 2.98 & 1\% compressive\\
\end{tabular}
\end{ruledtabular}
\end{table*}

We next track the structural change of MgO//Pt$_{100-x}$Al$_x$ based on the $x$-dependence of the XRD spectra plotted in Fig.~\ref{fig:XRD}a. The sputtered Pt film tends to assume fcc(111) texture on MgO(001) for a moderate $T_\textrm{sub}$, which is consistent with a previous work \cite{Gatel_Pt100}. Despite the lack of clear epitaxial relationship with the substrate, the strong fcc(111) diffraction peak and the clear Lau\'{e} fringes both imply the high crystallinity of the Pt film with sharp interfaces. Introducing Al changes the preferred orientation and texture of the alloy film to fcc(001) while drastically attenuates the Bragg diffraction of fcc(111). L1$_2$-ordered Pt$_{100-x}$Al$_x$ with no observable secondary phase was obtained for $x=26$. Further increase of $x$ leads to a rapid decrease of the out-of-plane lattice parameter $c$, as shown by the shift of the Bragg peaks towards higher $2\theta$ angles. After going through a fcc to bcc transition, single-phase B2-PtAl with clear Lau\'{e} fringes was eventually formed for $x=48$. In contrast, the XRD spectra of SiO$_2$//Pt$_{100-x}$Al$_x$ deposited at ambient temperature (Fig.~\ref{fig:XRD}b) show monotonic decrease of the fcc(111) peak intensity with increasing $x$, suggesting the non-equilibrium \cite{Masuda2020} fcc Pt-Al alloy gradually becomes amorphous upon incorporating more and more Al.

\begin{figure*}
\begin{center}
  \includegraphics[width=0.9\columnwidth]{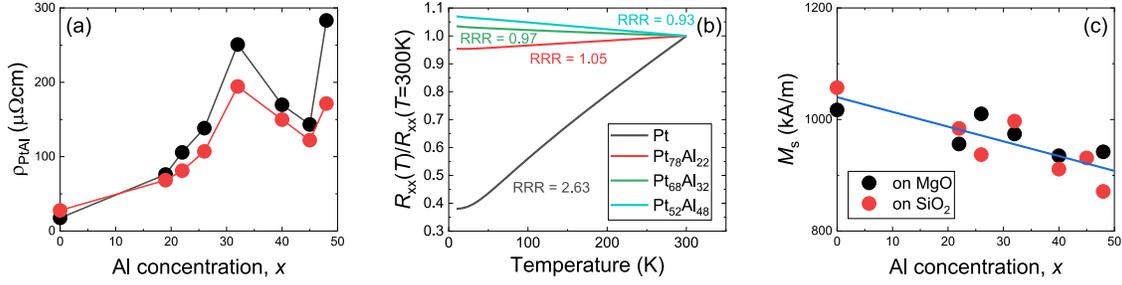}
  \caption{
  (a) $x$-dependence of longitudinal resistivity $\rho_\textrm{PtAl}$ for $\sim\SI{6}{\nano\meter}$ Pt$_{100-x}$Al$_x$ extracted from Pt$_{100-x}$Al$_x$($\sim$6)/CoFeB(2) bilayers. Data for samples on MgO are in black and those for samples on SiO$_2$ are in red. (b) Temperature dependence of longitudinal resistance $R_\textrm{xx}$ normalized by its value at \SI{300}{\kelvin} for Pt$_{100-x}$Al$_x$($\sim$6)/CoFeB(2) bilayers of selected Al compositions on MgO. The residual resistivity ratio (RRR) is defined as $R_\textrm{xx}(T=\SI{300}{\kelvin})$/$R_\textrm{xx}(T=\SI{10}{\kelvin})$. (c) $x$-dependence of saturation magnetization $M_\textrm{s}$. Blue line is a linear fit combining all the data.}
  \label{fig:DC_transport}
\end{center}
\end{figure*}

Focusing on MgO//Pt$_{74}$Al$_{26}$ and MgO//Pt$_{52}$Al$_{48}$, we measured the in-plane XRD ($2\theta\chi/\varphi$ scan; Fig.~\ref{fig:XRD}c and d) to reveal the tetragonal distortion of these compounds, which mainly arises from the substrate-induced in-plane biaxial strain. For MgO//Pt$_{52}$Al$_{48}$, we found an additional diffraction peak corresponding to bcc-PtAl(211) for the scan along MgO[100]. This parasitic orientation may partially explain the complicated RHEED patterns shown in Fig.~\ref{fig:RHEED}e and f. We further use $\varphi$ scans to probe the epitaxial relationship between these films and the substrates. We found Pt$_{74}$Al$_{26}$ film crystallizes in the cube-on-cube configuration on MgO, experiencing an in-plane tensile strain of $\sim ~7\%$ from the latter. Pt$_{52}$Al$_{48}$, upon making a \SI{45}{\degree} in-plane rotation, is facing an in-plane compressive strain of $\sim 1\%$ from the MgO. The density deduced from x-ray reflectivity, the in-plane ($a$) and out-of-plane ($c$) lattice parameters, and the tetragonal distortion ($c/a$) of these two compounds are summarized in Table~\ref{tab:XRD_summary}.

\subsection{\label{subsec:Transport}Basic magnetotransport and magnetic properties}
Magnetotransport was measured using the micro-fabricated Hall bar devices.
Figure~\ref{fig:DC_transport}a plots the $x$-dependence of the longitudinal resistivity $\rho_\textrm{PtAl}$ extracted from Pt$_{100-x}$Al$_{x}$/CoFeB bilayers using parallel circuit model and assuming $\rho_\textrm{CoFeB} = \SI{150}{\micro\ohm\centi\meter}$. In general, $\rho_\textrm{PtAl}$ for the two sample sets tend to increase with increasing $x$, reflecting the modulation of $\tau$ by tuning the alloy composition. $\rho_\textrm{PtAl}$ shows a pronounce local maximum for $x = 0.32$. We tentatively attribute this to the enhanced scattering due to the formation of many grain boundaries for this particular composition at the vicinity of fcc-bcc structural transition. Despite having higher crystallinity, Pt$_{100-x}$Al$_{x}$ layers grown on MgO tend to be more resistive than those grown on SiO$_2$. We consider the defects and dislocations due to epitaxial strain relaxation in these films can enhance the scattering.

The temperature $T$-dependence of $\rho_\textrm{PtAl}$ sheds light on the defects and impurities in these Pt$_{100-x}$Al$_{x}$ alloy films. Neglecting the $T$ dependence of $\rho_\textrm{CoFeB}$, we attribute the overall temperature change of the bilayer longitudinal resistance $R_{xx}$ to that of $\rho_\textrm{PtAl}$. The normalized $R_{xx}(T)/R_{xx}(T=\SI{300}{\kelvin})$ for selected samples on MgO are shown in Fig.~\ref{fig:DC_transport}b. We further define the residual resistivity ratio, $\mathrm{RRR}\equiv R_{xx}(T=\SI{300}{\kelvin})/R_{xx}(T=\SI{10}{\kelvin})$. On increasing $x$, RRR reduces and crosses 1 at $x\sim0.26$, reflecting a change from metallic-like (with a positive temperature coefficient of the resistivity) to semiconducting-like conduction (with a negative temperature coefficient of the resistivity). A dramatic enhancement of $\rho_\textrm{PtAl}$ for bcc-Pt$_{52}$Al$_{48}$ grown on MgO may reflect the intrinsic transport characteristics of this exotic metastable compound, although further optimization of the film growth is necessary to exclude the extrinsic transport contributions from the defects and impurities.

The Al concentration $x$-dependence of the saturation magnetization $M_\textrm{s}$ extracted from VSM for selected samples are plotted in Fig.~\ref{fig:DC_transport}c. Within experimental error (estimated to be $\sim 5\%$), no obvious substrate dependence of $M_\textrm{s}$ is observed. Instead, $M_\textrm{s}$ systematically decreases with increasing $x$, which may be related to the weakening of the proximity-induced magnetism at Pt$_{100-x}$Al$_x$/CoFeB interface, as Pt atoms at interface are gradually replaced by Al. Combining the two data sets, linear fit yields a slope of \SI{-264.1}{\kilo\ampere\per\meter\per\atomp} and a $y$-intercept of $M_\textrm{s,Pt} = \SI{1040}{\kilo\ampere\per\meter}$. Hereafter, we use the interpolated $M_\textrm{s}$ for the SOT quantification in these heterostructures.

\begin{figure}
\begin{center}
  \includegraphics[width=0.95\columnwidth]{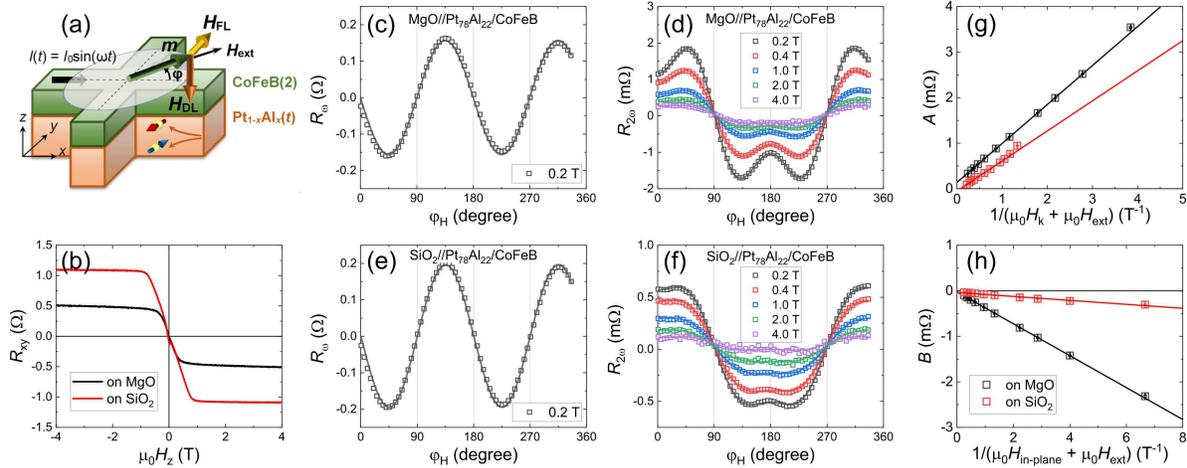}
  \caption{
  (a) Schematic illustration of the harmonic Hall measurement setup for quantifying the spin Hall efficiency of Pt$_{100-x}$Al$_x$/CoFeB bilayers with easy-plane magnetic anisotropy. An external magnetic field $\mu_0 H_\textrm{ext}$ is rotated within the azimuthal plane, making an angle $\varphi_\textrm{H}$ with the current $I$. (b) Hall resistance $R_{xy}$ against an out-of-plane external field $\mu_0 H_z$. (c) and (e) $\varphi_\textrm{H}$ dependence of the first harmonic Hall resistance $R_{\omega}$ with $\mu_0 H_\textrm{ext} = \SI{0.2}{\tesla}$. (d) and (f) $\varphi_\textrm{H}$ and $\mu_0 H_\textrm{ext}$ dependence of the second harmonic Hall resistance $R_{2\omega}$. (g) $\cos{\varphi_\textrm{H}}$ component ($A$) and (h) $\cos{2\varphi_\textrm{H}}\cos{\varphi_\textrm{H}}$ component of $R_{2\omega}$ plotted against $1/(\mu_0 H_\textrm{k}+\mu_0 H_\textrm{ext})$ and $1/(\mu_0 H_\textrm{in-plane}+\mu_0 H_\textrm{ext})$, respectively. All the data were measured at \SI{300}{\kelvin} for Pt$_{78}$Al$_{22}$(~6)/CoFeB(2) structures grown on MgO and SiO$_2$.}
  \label{fig:Harmonic}
\end{center}
\end{figure}

\subsection{\label{subsec:Harmonic}Harmonic Hall measurements}
\begin{figure*}
\begin{center}
  \includegraphics[width=0.9\columnwidth]{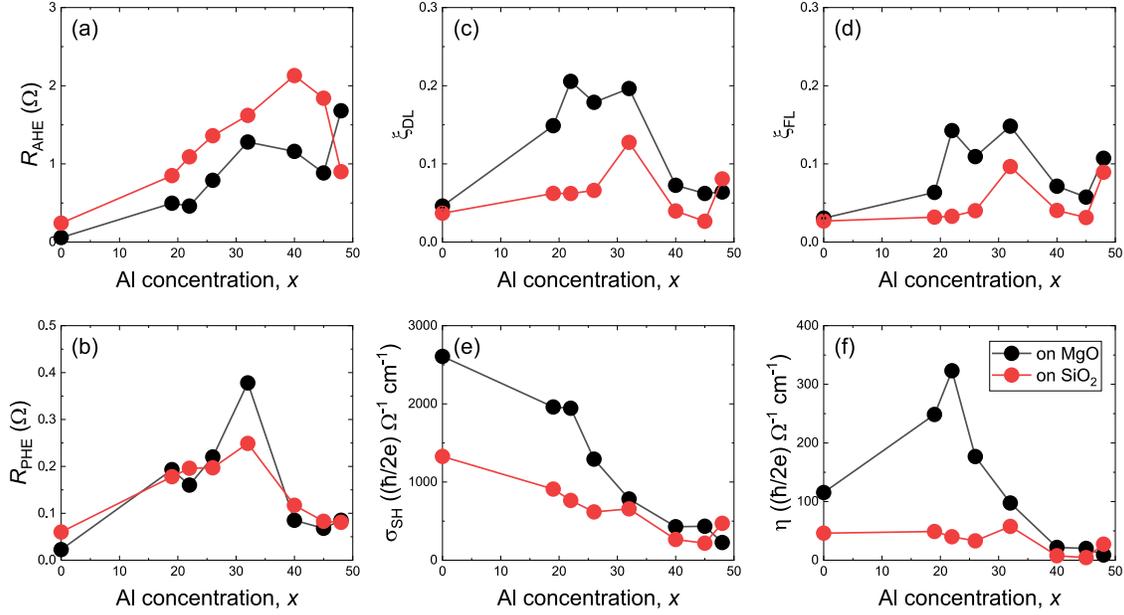}
  \caption{
  Al concentration $x$ dependence of (a) anomalous Hall resistance ($R_\textrm{AHE}$), (b) planar Hall resistance ($R_\textrm{PHE}$), (c) damping-like spin Hall efficiency ($\xi_\textrm{DL}$), (d) field-like spin Hall efficiency ($\xi_\textrm{FL}$), (e) spin Hall conductivity ($\sigma_\textrm{SH}$), and (f) the power efficiency $\eta$. Black and red symbols represent Pt$_{100-x}$Al$_x$/CoFeB bilayers grown on MgO and SiO$_2$ substrates, respectively.}
  \label{fig:Summary}
\end{center}
\end{figure*}
 The harmonic Hall technique \cite{KimNatMat2012_Harmonic,GarelloNatNano2013_SOT,HayashiPRB2014_Harmonic,AvciPRB2014_thermoelectric,ChiSciAdv2020_BiSb} was used to quantify the SOT in Pt$_{100-x}$Al$_{x}$/CoFeB bilayers with in-plane magnetization. The experimental setup is schematically depicted in Fig.~\ref{fig:Harmonic}a. An external field $\mu_0H_\textrm{ext}$ is applied in the PPMS while rotating the sample. The field vector is making a rotation within the sample azimuthal plane, making an angle $\varphi_\textrm{H}$ with the current, $I$. We measured the $\varphi_\textrm{H}$ angular dependence of the in-phase first harmonic ($R_\omega$) and out-of-phase second harmonic Hall resistances ($R_{2\omega}$) at various $\mu_0H_\textrm{ext}$ ranging from \SI{0.1}{\tesla} to \SI{4.0}{\tesla}. Including thermoelectric contributions, $R_\omega$ and $R_{2\omega}$ are of the form:
  \begin{eqnarray}
  \label{eq:R1w}
R_{\omega}=R_\mathrm{PHE}\sin2\varphi_\mathrm{H}
  \end{eqnarray}
 \begin{eqnarray}
\label{eq:R2w}
    R_{2\omega}=&\left(R_\mathrm{AHE}\dfrac{H_\mathrm{DL}}{H_\mathrm{ext}+H_\mathrm{k}}+R_\mathrm{const}\right)\cos\varphi_\mathrm{H}\\
    &-2R_\mathrm{PHE}\dfrac{H_\mathrm{FL}+H_\mathrm{Oe}}{H_\mathrm{ext}+H_\mathrm{in-plane}}\cos2\varphi_\mathrm{H}\cos\varphi_\mathrm{H} \nonumber\\
    &=A\cos\varphi_\mathrm{H}+B\cos2\varphi_\mathrm{H}\cos\varphi_\mathrm{H}\nonumber\\
    \nonumber
\end{eqnarray}
Harmonic Hall data collected for MgO//Pt$_{78}$Al$_{22}$(6)/CoFeB(2) and SiO$_2$//Pt$_{78}$Al$_{22}$(6)/CoFeB(2) bilayers are compared in Fig.~\ref{fig:Harmonic}b-h. The anomalous Hall resistance $R_\textrm{AHE}$ and the out-of-plane anisotropy field $\mu_0H_\textrm{k}$ are extracted from the anomalous Hall loop plotted in Fig.~\ref{fig:Harmonic}b. The extracted $\mu_0H_\textrm{k}$ are significantly lower than the demagnetizing field of CoFeB, indicating a non-negligible interface magnetic anisotropy contribution in Pt$_{100-x}$Al$_{x}$/CoFeB/Al trilayers. Eq.~\ref{eq:R1w} is used to fit the $\varphi_\mathrm{H}$-dependence of $R_\omega$ to extract the planar Hall resistance $R_\textrm{PHE}$, as shown in Fig.~\ref{fig:Harmonic}c and e. $\varphi_\textrm{H}$-dependence of $R_{2\omega}$ (Fig.~\ref{fig:Harmonic}d and f) can be decomposed into two parts according to Eq.~\ref{eq:R2w}. The prefactors $A$ (corresponding to $\cos\varphi_\mathrm{H}$ component) and $B$ (corresponding to $\cos2\varphi_\mathrm{H}\cos\varphi_\mathrm{H}$ component) of $R_{2\omega}$ are plotted against $1/(\mu_0 H_\textrm{k}+\mu_0 H_\textrm{ext})$ and $1/(\mu_0 H_\textrm{in-plane}+\mu_0 H_\textrm{ext})$, respectively. We assume an in-plane anisotropy field $\mu_0 H_\textrm{in-plane} = \SI{0.05}{\milli\tesla}$ which mainly arises from the four-fold symmetry of the Pt$_{100-x}$Al$_x$. The slopes of the linear fits in Fig.~\ref{fig:Harmonic}g and Fig.~\ref{fig:Harmonic}h were used to extract the damping-like and the field-like spin-orbit effective fields. The corresponding spin Hall efficiencies $\xi_\textrm{DL}$ and $\xi_\textrm{FL}$ are estimated based on the following equation:
\begin{eqnarray}
\label{eq:xi_DL}
\xi_\mathrm{DL(FL)} = \frac{2e}{\hbar} \frac{\mu_0 H_\mathrm{DL(FL)} M_\mathrm{s} t_\mathrm{CoFeB}}{j_\mathrm{PtAl}}
\end{eqnarray}
where $M_\mathrm{s}$ is the interpolated saturation magnetization of the CoFeB layer obtained from Fig.~\ref{fig:DC_transport}c. Knowing the resistivity of each layer, we can now calculate the spin Hall conductivity $\sigma_\textrm{SH}$ and the power efficiency $\eta$. Similar analyses were repeated for the two series of $x$-varying samples grown on MgO and SiO$_2$ to establish the substrate and $x$-dependence of these quantities. Results are summarized in Fig.~\ref{fig:Summary}.

$R_\textrm{AHE}$ and $R_\textrm{PHE}$ are plotted against $x$ in Fig.~\ref{fig:Summary}a and b, respectively. The anomalous Hall loops of all the samples can be found in the Supplementary Material. Although Pt$_{100-x}$Al$_{x}$ films grown on SiO$_2$ are typically more conducting and should shunt more current from the CoFeB, we surprisingly found higher $R_\textrm{AHE}$ for SiO$_2$//Pt$_{100-x}$Al$_{x}$/CoFeB series. Instead, $x$-dependence of $R_\textrm{PHE}$ for the two sample series are quite similar. This can be understood because CoFeB has a relatively small anisotropic magnetoresistance and the planar Hall effect in Pt$_{100-x}$Al$_{x}$/CoFeB should mainly originate from the spin Hall magnetoresistance (SMR) \cite{Nakayama2013,Chen2013,Kim2016}. Details on the SMR of these samples are elaborated in the Supplementary Material.

The primary results of this work are shown in Fig.~\ref{fig:Summary}c-f. For $x \lesssim 30$, we found striking enhancement for structures on MgO over structures on SiO$_2$ across all aspects, including $\xi_\mathrm{DL}$ (Fig.~\ref{fig:Summary}c), $\xi_\mathrm{FL}$ (Fig.~\ref{fig:Summary}d), $\sigma_\textrm{SH}$ (Fig.~\ref{fig:Summary}e), and $\eta$ (Fig.~\ref{fig:Summary}f). For example, for Pt$_{78}$Al$_{22}$/CoFeB bilayer on MgO, with $\rho_\textrm{PtAl} = \SI{105}{\micro\ohm\centi\meter}$, we found $\xi_\textrm{DL} \sim +0.20$ and $\sigma_\textrm{SH} \sim 1900 (\hbar/2e)\Omega^{-1}\text{cm}^{-1}$, whereas we only obtained $\xi_\textrm{DL} \sim +0.06$ and $\sigma_\textrm{SH} \sim 770 (\hbar/2e)\Omega^{-1}\text{cm}^{-1}$ for the structure of the same $x$ on SiO$_2$ with a lower degree of crystallinity and exhibiting a slightly lower $\rho_\textrm{PtAl} = \SI{81}{\micro\ohm\centi\meter}$. This represents up to seven-fold increase in $\eta$, hence a seven-fold reduction of power consumption is expected. Compared to a previous work \cite{PtAl_APL2016}, similar $\xi_\textrm{DL}$ and $\sigma_\textrm{SH}$ can only be achieved in Pt$_{85}$Hf$_{15}$(5.5)/Pt(0.5)/Co trilayer with carefully engineered interface. We may thus expect further enhancement of $\sigma_\textrm{SH}$ and $\eta$ for well-ordered Pt$_{78}$Al$_{22}$/CoFeB upon optimizing the interface.

The strongly fcc(111)-textured Pt (i.e. $x = 0$) on MgO also exhibits a relatively large $\sigma_\textrm{SH} \sim 2600 (\hbar/2e)\Omega^{-1}\text{cm}^{-1}$ with a low $\rho_\textrm{Pt} = \SI{17.6}{\micro\ohm\centi\meter}$. This is to be compared with the relatively poorly crystallized, also fcc(111)-textured Pt on SiO$_2$, with $\sigma_\textrm{SH} \sim 1300 (\hbar/2e)\Omega^{-1}\text{cm}^{-1}$ and $\rho_\textrm{Pt} = \SI{27.8}{\micro\ohm\centi\meter}$. Here, the effect of crystallinity may also present. However, in view of the low $\rho_\textrm{Pt}$, additional $\sigma_\textrm{SH}$ contribution due to extrinsic skew scattering mechanism cannot be excluded. Careful evaluation of SOT at various temperatures may allow separation of the two $\sigma_\textrm{SH}$ contributions. Finally, the rapid decrease of the charge-to-spin conversion efficiency in bcc Pt$_{100-x}$Al$_{x}$ with $x > 32$ suggests for Pt-rich alloys the fcc crystal structure and $\sigma_\textrm{SH}$ are strongly correlated. The high crystallinity alone is a necessary but not a sufficient condition for obtaining large $\sigma_\textrm{SH}$.

We also found large positive $\xi_\mathrm{FL}$ with a maximum at $x=0.22$, which exhibits similar $x$-dependence as $\xi_\mathrm{DL}$. The observed $\xi_\mathrm{FL}$ is opposing the Oersted field with a $\xi_\mathrm{FL}/\xi_\mathrm{DL}$ ratio up to 0.7, suggesting the two quantities are strongly correlated. We consider the bulk spin Hall effect contribution to $\xi_\mathrm{FL}$ scales with $\xi_\mathrm{DL}$ and is responsible for the large $\xi_\mathrm{FL}$ in Pt-Al/CoFeB bilayers with $\xi_\mathrm{DL}$ up to 0.20. In contrast, the $\xi_\mathrm{FL}$ contribution from the interfacial Rashba-Edelstein effect should be negligible in these more resistive heterostructures because of its small size of $\approx 0.02$, which does not scale with $\rho_\textrm{PtAl}$ \cite{DuPRA2020}.

\begin{figure*}
\begin{center}
  \includegraphics[width=0.9\columnwidth]{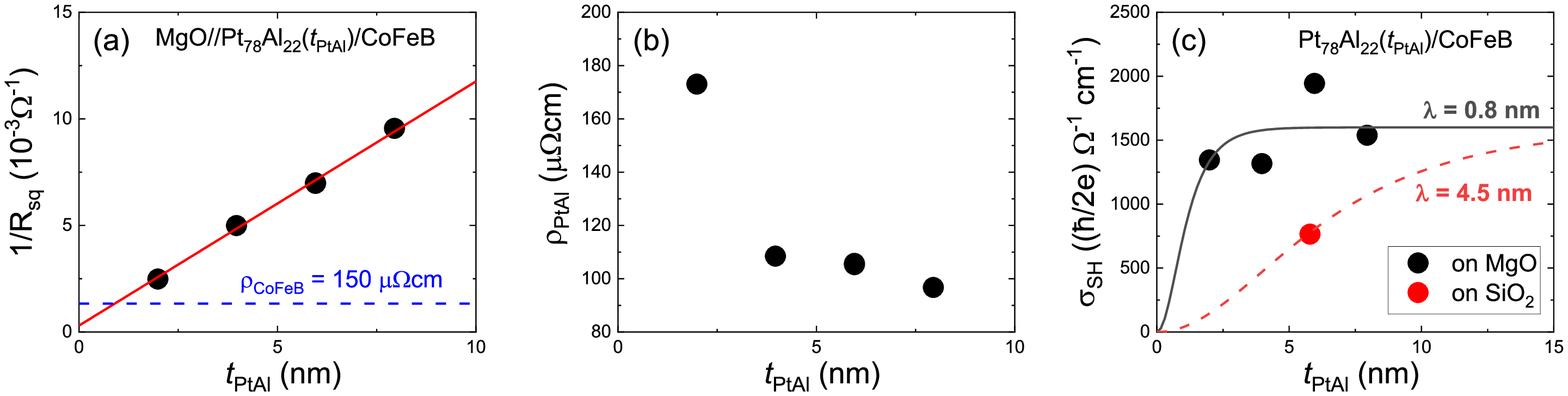}
  \caption{
  (a) Sheet conductance $1/R_\textrm{sq}$ against Pt$_{78}$Al$_{22}$ layer thickness $t_\textrm{PtAl}$ for Pt$_{78}$Al$_{22}$($t_\textrm{PtAl}$)/CoFeB bilayers. Blue horizontal dashed line represents the expected sheet conductance of the CoFeB layer. (b) Resistivity of Pt$_{78}$Al$_{22}$ layer $\rho_\textrm{PtAl}$ assuming a fixed CoFeB resistivity $\rho_\textrm{CoFeB} = \SI{150}{\micro\ohm\centi\meter}$. (c) $t_\textrm{PtAl}$ dependence of spin Hall conductivity $\sigma_\textrm{SH}$. Black and red curves are calculated from Eq.~\ref{eq:sigma_SH} using the same $\overline{\sigma}_\mathrm{SH}$ in the bulk limit and different spin diffusion length $\lambda$.}
  \label{fig:thickness}
\end{center}
\end{figure*}

We have also investigated the PtAl thickness $t_\textrm{PtAl}$ dependence of SOT for Pt$_{78}$Al$_{22}$($t_\textrm{PtAl}$)/CoFeB structures grown on MgO, which exhibits the highest $\eta$. The sheet conductance $1/R_\textrm{sq}$ against $t_\textrm{PtAl}$ is plotted in Fig.~\ref{fig:thickness}a. Assuming a fixed CoFeB resistivity ($\rho_\textrm{CoFeB} = \SI{150}{\micro\ohm\centi\meter}$), the extracted resistivity of PtAl $\rho_\textrm{PtAl}$ is shown in Fig.~\ref{fig:thickness}b. The strong enhancement of $\rho_\textrm{PtAl}$ for low $t_\textrm{PtAl}$ may be attributed to interfacial scattering and defects in ultrathin films. Since the intrinsic mechanism is expected to dominate, we focus on the $t_\textrm{PtAl}$ dependence of $\sigma_\textrm{SH}$ as shown in Fig.~\ref{fig:thickness}c. For the structure on MgO with $t_\textrm{PtAl}$ as thin as $\sim\SI{2}{\nano\meter}$, we found a relatively large $\sigma_\textrm{SH} \sim 1350 (\hbar/2e)\Omega^{-1}\text{cm}^{-1}$. The fact that all the four data points for structures on MgO clearly exhibit higher $\sigma_\textrm{SH}$ than the structure of the same $x$ grown on SiO$_2$ (red symbol) confirms the robustness of our results.
Within a drift-diffusion model, for $t_\textrm{PtAl}$ comparable to the spin diffusion length $\lambda$, $\sigma_\textrm{SH}$ can be described by \cite{NguyenPRL2016}:
\begin{eqnarray}
\label{eq:sigma_SH}
\sigma_\mathrm{SH}(t_\mathrm{PtAl}) = \overline{\sigma}_\mathrm{SH} \left[ 1-\mathrm{sech} {\frac{t_\mathrm{PtAl}}{\lambda}} \right]
\end{eqnarray}
Black curve in Fig.~\ref{fig:thickness}c is calculated using Eq.~\ref{eq:sigma_SH} with $\overline{\sigma}_\mathrm{SH} = 1600 (\hbar/2e)\Omega^{-1}\text{cm}^{-1}$ and $\lambda = \SI{0.8}{\nano\meter}$. The thickness dependence of the SMR is also consistent with such a short $\lambda$ (See the supplementary material). The red dashed curve shows that one needs to assume a much longer $\lambda \sim \SI{4.5}{\nano\meter}$ for the structure on SiO$_2$ to account for the $\sigma_\textrm{SH}$ gap between the two sample sets. If the spin relaxation is governed by the Elliott-Yafet mechanism \cite{Elliott,Yafet}, $\lambda$ should be inversely proportional to $\rho_\textrm{PtAl}$. Yet, \SI{2}{\nano\meter} Pt$_{78}$Al$_{22}$ layer on MgO with the highest $\rho_\textrm{PtAl}$ is only about twice more resistive than Pt$_{78}$Al$_{22}$ layer grown on SiO$_2$.
Therefore, we conclude that the potential change of $\lambda$ alone cannot explain the dramatic enhancement of $\xi_\textrm{DL}$ and $\sigma_\textrm{SH}$ for highly fcc-textured Pt-Al alloys.

\section{\label{sec:Discussion}Discussion}
First-principles calculations suggest the exceptionally high intrinsic $\sigma_\mathrm{SH}$ of fcc-Pt is mainly due to the double degeneracies near the Fermi level $E_\textrm{F}$ at the high-symmetry $L$ and $X$ points of the fcc lattice \cite{Guo_PRL_PtSHC}. Notably, $E_\textrm{F}$ of fcc-Pt falls practically on the summit of a high $\sigma_\textrm{SH}$ peak with a broad full-width at half maximum (FWHM) of the order of \SI{1}{\electronvolt}. Although this SHE contribution is robust against impurities due to its intrinsic nature, introducing dopants unavoidably modulate $\sigma_\mathrm{SH}$ via three mechanisms. Firstly, alloying may induce carrier doping that shifts the $E_\textrm{F}$ of the doped Pt alloy away from this optimum position, leading to a reduction of $\sigma_\textrm{SH}$. Secondly, alloying with a lighter element will weaken the average spin-orbit coupling and distort the critical nearly-degenerated nodes/lines in the band structure. Thirdly, alloying can deform the lattice and even induce structural change that will significantly alter the band structure. We note that rigid-band approximations may be valid for the first mechanism whereas the later two are clearly beyond this simple picture. Based on these three mechanisms, we can now comment on our experimental observations.

In Fig.~\ref{fig:Summary}e, we have observed a monotonic decrease of $\sigma_\textrm{SH}$ with increasing $x$. In principle, all the three mechanisms may be relevant and are hardly distinguishable. Here, the comparison of $\sigma_\textrm{SH}$ for fcc-Pt$_{100-x}$Al$_{x}$ (grown on different substrates) exhibiting different degree of crystallinity allows us to extract the net $\sigma_\textrm{SH}$ gain which is correlated with the high-quality fcc texture of the L1$_2$-ordered Pt$_3$Al alloy. In such an ordered alloy, the Al atoms selectively substitute Pt atoms of the parent fcc-Pt lattice which helps to maintain the fcc structure throughout and minimize the band structure distortion. Instead, if the same amount of Al is randomly distributed in Pt (the film grown on SiO$_2$), due to their small atomic radius, Al atoms are more likely occupying interstitial sites, resulting in a rapid degradation of the fcc structure. The unique degenerated bands with large spin Berry curvature in the electronic band structure of fcc-Pt may be strongly perturbed and one may expect a faster reduction of $\sigma_\textrm{SH}$ in this case. Having established the tunability of $\sigma_\textrm{SH}$ with the degree of crystallinity, it would be interesting to check whether the performance of other Pt-based fcc alloys (e.g. Pt-Au, Pt-Pd and Pt-Cu) can be further enhanced by improving the film texture or atomic ordering.

Furthermore, we note that fully epitaxial films are not required for observing such an enhancement as demonstrated by MgO//Pt$_{78}$Al$_{22}$/CoFeB which, according to the XRD spectrum, is consisted of a mixture of fcc(111) and fcc(100) textured crystallites. We believe this is partially due to the fact that the measured spin current is flowing along the film normal. In this geometry, the long-range atomic order within the film plane is less critical. We should also note that the high crystallinity alone is not a sufficient condition for obtaining large $\sigma_\textrm{SH}$, as B2 Pt$_{52}$Al$_{48}$ with relatively high crystallinity only exhibit low $\sigma_\textrm{SH}$. The band structure of bcc-based Pt compounds may not possess nearly-degenerated nodes/lines near $E_\textrm{F}$ that can host large spin Berry curvature.

Finally, from the applications point of view, protocols for growing at ambient temperature L1$_2$ Pt$_3$Al (or other fcc-based Pt compounds) of similar quality on amorphous or polycrystalline substrates are highly desirable. Inspired by the successful growth of L1$_0$-ordered FePt \cite{Yasushi2001,Xu2004} and tetragonal Mn-based Heusler compounds \cite{Lee2015,Jeong2016} on SiO$_2$, we believe some of these strategies may be applied to L1$_2$ Pt$_3$Al. Furthermore, the fcc(100) texture of L1$_2$ Pt$_3$Al with four-fold in-plane symmetry is compatible with the coherent tunneling across the crystalline MgO(001) barrier, a prerequisite for large tunneling magnetoresistance readout \cite{Yuasa2007}.

\section{\label{sec:Summary}Summary}
In summary, we have studied the Al concentration $x$ dependence of spin Hall efficiency and power consumption efficiency for Pt$_{100-x}$Al$_{x}$/CoFeB bilayers grown on two different substrates (MgO or SiO$_2$). For $x \lesssim 30$, systematic enhancement over all aspects was found for structures grown on MgO, which we attribute to the ideal positioning of Al atoms substituting Pt in the fcc lattice. The fcc crystal structure critical for large spin Berry curvature can be maintained with better crystallinity and higher ordering compared to similar structures grown on SiO$_2$. This work uncovers the essential role of the structural aspect for modulating the spin Hall efficiency in Pt-based fcc alloys, pointing to a new direction for further improving state-of-the-art materials for spin current generation via the spin Hall effect.

\section*{Supplementary Material}
See supplementary material for the comparison of x-ray reflectivity, spin Hall magnetoresistance, and anomalous Hall effect for samples grown on MgO and SiO$_2$ substrates.

\begin{acknowledgments}
This work was supported by JSPS KAKENHI Grant-in-Aid for Scientific Research (S) (JP18H05246), Grant-in-Aid for Early-Career Scientists (JP20K15156) and Grant-in-Aid for Scientific Research (A) (JP20H00299). The structural characterizations and device fabrication were carried out at the Cooperative Research and Development Center for Advanced Materials, IMR, Tohoku University. The authors thank T. Sasaki for her help to do the film deposition by ion beam sputtering.
\end{acknowledgments}

\section*{Data Availability Statement}
The data that support the findings of this study are available from the corresponding author upon reasonable request.

\nocite{*}
%

\end{document}